# The Evolution of Agile and Hybrid Project Management Methodologies: A Systematic Literature Review

BIANCA, B, LEECH

Academy of Computer Science and Software Engineering, University of Johannesburg, Johannesburg, Gauteng, South Africa, biancaleech29@gmail.com

RIDEWAAN, R, HANSLO

Academy of Computer Science and Software Engineering, University of Johannesburg, Johannesburg, Gauteng, South Africa, ridewaanh@uj.ac.za

The rapid evolution of IT projects has driven the transformation of project management methodologies, from traditional waterfall approaches to agile frameworks and, more recently, hybrid models. This systematic literature review investigates the evolution of agile methodologies into hybrid frameworks, analysing their implementation challenges and success factors. We identify key trends through PRISMA-guided analysis of peer-reviewed studies from the last 8 years. Hybrid methodologies emerge from agile limitations in large-scale and regulated environments, combining iterative flexibility with structured governance. Agile has several implementation challenges, leading to hybrid methods, and the success hinges on leadership support, tailored process integration, and continuous improvement mechanisms. The study explores the need for contextual adaptation over rigid frameworks, offering practical insights for organisations navigating hybrid transitions.

**CCS CONCEPTS** • Software and its engineering • Software creation and management • Software development process management

**Additional Keywords and Phrases:** Agile, Hybrid, Project Management, Methodology Evolution

## 1 INTRODUCTION

The rapid evolution of technology and increasing demand for IT projects have driven advancements in project management methodologies to ensure project success. Traditional approaches like waterfall are structured but often inadequate for dynamic, fast-paced environments. This led to the widespread adoption of agile methodologies in software development and the IT sectors. Agile methods, emphasising flexibility, iterative progress, and customer collaboration, have significantly improved project performance, particularly in dynamic environments [1].

While agile methodologies have transformed project management, their limitations in large-scale and regulated environments have led to the emergence of hybrid approaches. Studies have tested agile methodologies with diverse stakeholders in complex projects and at the enterprise level, revealing challenges driving hybrid frameworks' evolution, combining agile flexibility with traditional structure and predictability [2].

This Systematic Literature Review (SLR) examines this evolutionary trajectory, focusing on two key aspects: (1) the transition from pure agile to hybrid methodologies and (2) implementation challenges and success factors. We analysed numerous studies to identify patterns, trends, and best practices in modern project management, leveraging insights gathered through the PRISMA flowchart. The paper is structured as follows: Section 1 introduces the topic. Section 2 details the research methodology, including data collection and filtering. Section 3 presents the search results and quality evaluation. Section 4 discusses the thematic analysis and findings. Finally, Section 5 concludes with implications and future research directions.

## 2 RESEARCH METHOD

The research method followed for this SLR aims to identify, investigate, and incorporate existing official, published research to answer a formulated research question. Data was collected by querying databases using search terms and filters to identify relevant papers. The method addresses key literature review steps: identifying goals, defining methods, collecting data, collating literature, extracting features, and analysing the results [3].

### 2.1 Research questions and objective

The objective is to analyse the evolution from pure agile to hybrid methodologies and their implementation challenges and success factors.
1. How have agile methodologies evolved into hybrid frameworks?
2. What are the key challenges and success factors in implementing hybrid approaches?

### 2.2 Search terms

The following search strings were used:
("agile" OR "kanban" OR "scrum" OR "XP" OR "extreme programming" OR "scrumban" OR "learn development") AND ("hybrid" OR "water-scrum-fall" OR "agile-stage-gate" OR "v-model" OR "tailoring" OR "method engineering") AND ("project management" OR "methodology evolution" OR "framework*" OR "integration") AND ("challenge*" OR "success factor*" OR "implementation" OR "scaling" OR "adaptation").

### 2.3 Selection Criteria

Table 1 below summarises the inclusion and exclusion criteria per category.

Table 1 Inclusion-exclusion criteria

| Category | Inclusion Criteria | Exclusion Criteria |
| --- | --- | --- |
| Methodology Focus | Examines Agile or hybrid methods | Focuses more on traditional Project management |
| Language | English | Any other language than English |
| Publication Type | Journal Articles, Literature reviews, Conferences | Non-peer-reviewed work (e.g., blogs, editorials, predatory journals), Conference abstracts without full papers |
| Publication year | Last eight years (2017 - 2025) | Published before 2017 |
| Accessibility | Full text available | Duplicate papers, Full text not available |

### 2.4 Source Selection

The main databases searched include Google Scholar, Scopus, Elsevier, IEEE Xplore, and Semantic Scholar. These databases contain several papers related to project management and are therefore suitable databases for this review. These databases have free access provided by the institution, making it convenient for research.

### 2.5 Prisma Flowchart

Figure 1 depicts the four stages of the Prisma flowchart, with 21 studies included as part of the SLR synthesis.

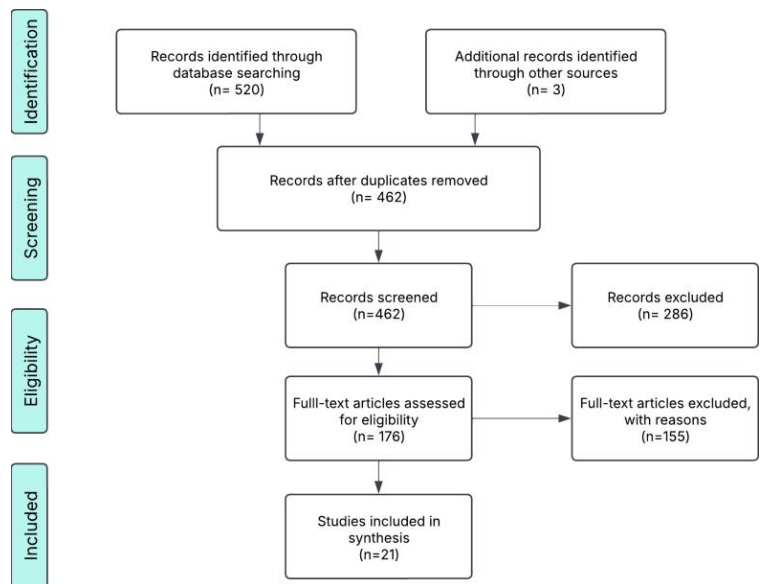

Figure 1 Prisma Flowchart

## 2.6 Data Extraction

Thematic Analysis (TA) detects, examines, and interprets meaningful patterns, or "themes" in qualitative data. TA functions as a flexible tool or procedure, not tied to any theoretical frameworks, rather than a methodology, which is a research approach grounded in and limited by theoretical perspectives [5]. By analysing overlapping patterns and commonalities in the data, key themes were identified to address the research questions of this SLR. The data extraction table, provided in the following link: https://bit.ly/SLR_appendices, summarises the details of each study, including its citation, source, database origin, and key concepts derived from thematic analysis. The concepts column lists the specific codes, while the main concept column highlights the overarching themes. The analysis followed a structured coding approach to enhance methodological rigour. The primary researcher independently performed open coding on the selected studies to identify initial concepts, followed by axial coding to group these into broader themes.

## 3 RESULTS

### 3.1 Search Results

The search methodology outlined in Sections 2.2 to 2.5 included 21 studies for the synthesis in this paper. Figure 2 illustrates key concepts identified from the studies, and Figure 3 shows the distribution of included studies by publication year, revealing that the highest number of papers (9) appeared in 2024. Figure 4 displays the distribution of articles retrieved from each database, following the source selection criteria in Section 2.4. As mentioned earlier, the structured coding approach of open coding followed by axial coding resulted in a conceptual matrix mapping challenges, success factors, and evolutionary patterns across different project contexts [25, 26]. This process aligns with PRISMA guidelines for systematic reviews by ensuring transparent and reproducible analysis methods. Figure 5 maps the Agile to hybrid evolution, depicting the Agile limitations and the hybrid challenges and success factors. These results are discussed in Section 4.

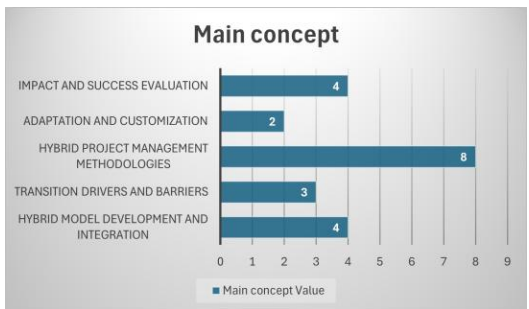
Figure 2 Main concepts identified

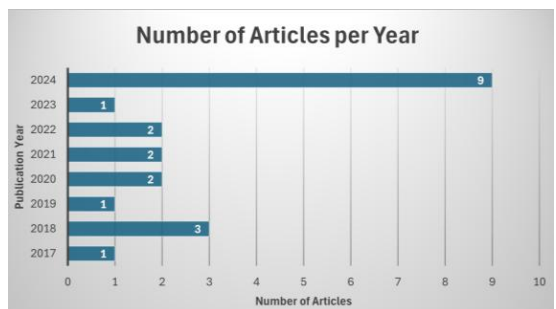
Figure 3 Articles included in synthesis

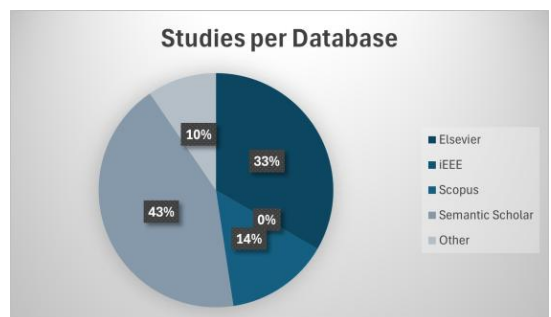
Figure 4 Number of articles per database

### 3.2 Quality evaluation of articles

Due to the papers having gone through inclusion and exclusion criteria and the PRISMA chart, the articles are very applicable for the review. The quality evaluation definition and results provided in the following link: https://bit.ly/SLR_appendices indicate that 16 of the documents have a score above 4.5, where 10 papers have a score of 5. Thus, the papers included can contribute immensely to the review, as they are relevant [4].

## 4 DISCUSSION

The evolution from Agile to hybrid methodologies emphasises the dynamic nature of project management as it adapts to increasing complexity, regulatory demands, and scalability challenges in projects. This section synthesises the findings from the reviewed literature to address the research. As organisations increasingly adopt project-driven frameworks, their approaches to delivering value have shifted accordingly. Project practices and methodologies evolve and adapt dynamically to changing environments to meet the organisation's needs [6]. Traditional methods are too rigid for dynamic markets, resulting in inefficiencies in adapting to the evolving requirements. The increase in software integration into physical products further highlighted the need for Agile's iterative flexibility [7]. However, Agile also has limitations, leading to the development of Hybrid models, which combine Agile's flexibility with traditional structures to manage projects requiring both traits [8]. 52% of projects adopt hybrid approaches, reflecting a trend toward blending methodologies [9].

### 4.1 Agile Methodologies

Large-scale agile projects involve implementing agile principles and practices across multiple teams collaborating on complex, often enterprise-wide initiatives. This requires synchronising and aligning numerous agile teams to achieve cohesive delivery, effective teamwork, and seamless integration. Scaling agile is essential for organisations that sustain a competitive edge in dynamic and complex market environments. As companies grow, their operations, projects, and product portfolios typically increase in scale and complexity. Initially designed for small, co-located teams, agile methodologies must be adapted to larger contexts to retain their core benefits, such as adaptability, rapid delivery, and responsiveness across the entire enterprise [2].

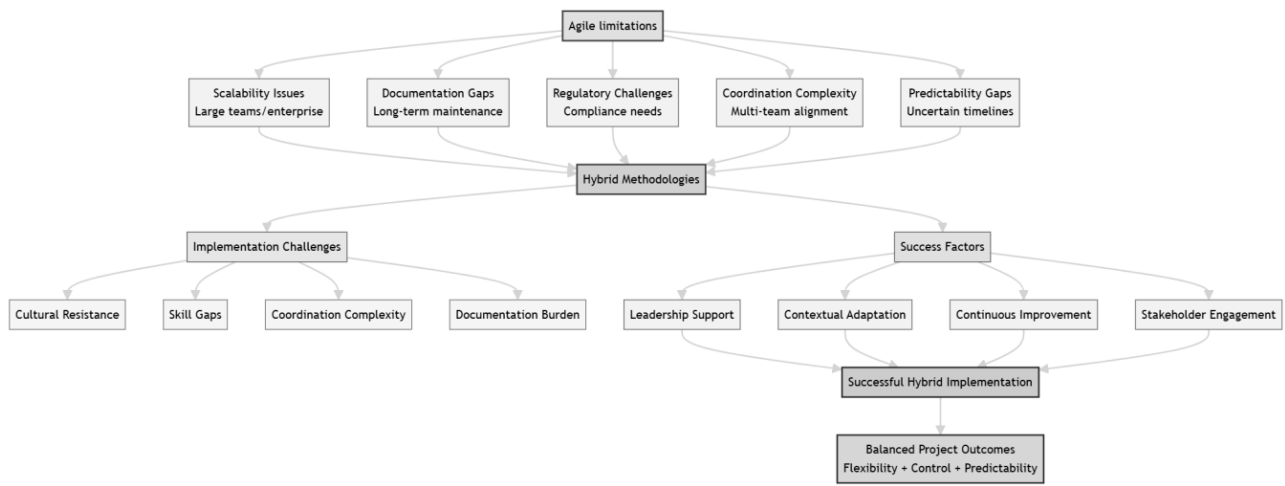

Figure 5 Conceptual framework mapping the Agile to hybrid evolution

Agile frameworks are not universally suitable, as not all Agile methodologies cater for the organisation's needs and cultural adaptation. Hence, many organisations struggle to adapt to their needs, emphasising the need for hybrid methods [2]. Agile's limitations in addressing risks like proper utilisation of funds or complex tasks highlight the necessity for structured adaptations [11]. Successful implementation of Agile requires a lot of commitment, communication, and a culture that is receptive to change and new challenges [6]. For example, convincing customers with alternative technologies often requires blending Agile's flexibility with traditional stakeholder management practices [11]. Hybrid contexts also necessitate redefining traditional critical success factors (CSF), which include team collaboration and customer involvement [12]. Traditional agile frameworks face additional challenges. Scrum's product-centric focus limits its adaptability across project types. It also struggles with remote collaboration, large teams, and multi-team coordination, creating gaps in team dynamics and customer interaction [12,13,17].

Agile's focus on rapid delivery and adaptability can deteriorate design quality. Testing may be deprioritised to accommodate changing requirements, leading to increased fault tolerance. Agile's emphasis on prioritising software over comprehensive documentation risks knowledge gaps in long-term maintenance [7,14]. Effort and timeline estimations in Agile are unpredictable due to dynamic factors like team changes and new tools [14]. Scope creep, lack of upfront planning, and documentation-heavy environments are significant Agile challenges [9,15], particularly in distributed teams where clients demand documentation and accountability [8,16]. Agile lacks mechanisms for strategic project selection or termination [7]. Traditional management's rigidity contrasts with Agile's flexibility but offers stability that Agile may lack [18]. Maintaining dedicated teams in challenging modern development contexts like manufacturing is another challenge [7].

## 4.2 Evolution from Agile to Hybrid Methodologies

Hybrid methodologies represent a mature evolution rather than a transitional phase, validated by their effectiveness in achieving stakeholder success comparable to pure Agile [9]. As organisations scaled Agile beyond small teams, limitations became evident. Agile's neglect of documentation and challenges in managing physical product iterations hindered long-term maintenance and compliance, further driving the need for hybrid models [19]. Risks like "difficulty in adjusting schedules to customer timelines" and "imperfect sprint planning", identified through the Analytic Hierarchy Process (AHP), reveal gaps in pure Agile frameworks [11]. Hybrid methodologies emerged to combine Agile's flexibility with traditional methods' predictability, addressing the need for adaptability in complex projects [11,15,16]. Hybrid frameworks like Scaled Agile Framework (SAFe), Scrum-Stage-Gate, and Modified Agile for Hardware Development (MAHD) bridge these gaps [10,19].

Large-scale IT projects require cross-team alignment, which pure agile frameworks like Scrum struggle to achieve. Hybrid adoption is rising, with 20% of organisations blending Agile and plan-driven methods to address scalability, complexity, and organisational policy alignment [6]. IBM's "Agile with Discipline" model exemplifies this evolution by balancing adaptability with traditional, structured documentation and planning to meet clients' needs [16]. Successful hybrid adoption requires digital leadership to navigate physical and digital collaboration spaces. For example, xFarm Technologies uses agile-stage-gate hybrids, where leaders orchestrate in-person ideation sessions for strategic alignment, while digital tools are used for asynchronous execution tracking and communication. This addressed many coordination challenges in distributed teams [20].

Performance and scalability issues are key drivers of Agile's evolution, as these methodologies struggle at enterprise levels [13,21]. New frameworks like SAFe and Large-scale Scrum (LeSS) were developed to address the scalability issues, but they introduced complexity and rigidity, which counteract Agile's flexibility. Hybrid approaches like Scrumban combine agile practices with traditional methods to manage large-scale projects better [2]. Agile teams prioritise informal channels for 73% of the execution-phase interactions, whereas Waterfall teams rely on formal ticketing systems for documentation. Hybrid setups often enable faster decision-making than pure Agile, mitigating scalability issues [22]. Industries like healthcare and finance require flexibility, structured documentation, and compliance, which pure Agile lacks [1]. Hybrid models integrate Agile's iterations with plan-driven documentation to ensure adherence to regulations without sacrificing agility [23], particularly when combined with DevOps practices for continuous compliance [21]. Hybrid models like MAHD enforce iterative validation cycles and modular design to meet regulatory standards without sacrificing agility [10].

Frameworks like Agile-Stage-Gate combine Scrum's iterative cycles with staged phases of the Stage-Gate process [7,23], while integrating Cloud technologies with Agile and DevOps enables scalable hybrid workflows [21]. Examples of Hybrid methodologies include Water-Scrum-Fall, which segments projects into planning using Waterfall, iterative development through Scrum, and delivery through Waterfall phases. This retains flexibility while adding structure [10,19,23]. Organisations tailor methodologies based on factors like project type, team size, and maturity. They also tailor hybrids to project needs, integrating Scrum with Rational Unified Process (RUP) for large-scale projects and combining Lean-Agile principles for efficiency [6]. SAFe reduces time-to-market and increases productivity by 75%. It also improves quality and employee satisfaction by 50% [13]. Agile-Waterfall Hybrid and Scrum-Waterfall Hybrid illustrate structured integration strategies [15].

Water-Scrum-Fall mitigates Scrum's limitations by incorporating structured planning and delivery phases, enhancing Scrum's scalability [23], especially when combined with cloud-based continuous integration and continuous deployment (DevOps) pipelines [21]. Additional Hybrids like Kanban-Gantt merge Kanban's visual workflows with Gantt charts' timeline tracking to ensure better time-tracking in complex projects. This combination enhances visibility and coordination in complex projects but may increase documentation overhead [2]. Agile-Stage-Gate combines Agile's feedback loops with Stage-Gate's milestone-driven control, balancing responsiveness and governance, offering faster feedback loops than the traditional waterfall while maintaining structured points [7, 23]. This hybrid model improves responsiveness without sacrificing control [7], like how DevOps enhances Agile's continuous delivery capabilities [21]. Additionally, the Agile-Stage-Gate offers a middle ground between agile and waterfall for industries like manufacturing [7].

### 4.3 Implementation challenges in Hybrid methodologies

Despite all the benefits of adopting Hybrid methodologies, they face multiple challenges and adoption barriers. These challenges include cultural resistance, coordination complexity, skill gaps, and documentation burden. Additional challenges include team alignment, empirically defining hybrid approaches, and managing practice combinations without clear theoretical frameworks [9,15]. Balancing agile flexibility with traditional documentation and managing dispersed teams are also identified as challenges [16]. Legacy processes and hierarchical structures often clash with Agile's collaborative ethos. The risk factor "adaptation of Agile methodology for the first time" highlights resistance to new practices, which hybrid models must address through phased integration [11]. The middle management may resist hybrid models due to perceived loss of control [2]. Cultural resistance is one of the most significant barriers, as teams accustomed to pure Agile may resist reintroducing traditional processes. They don't want to or are afraid to learn new things [13,21].

Some organisations' teams have inconveniences when making changes to project plans, because they were unsure about how the contingency plans worked, causing a learning curve. Due to the lack of training and knowledge, there were project delays up to three months [17]. One's mindset plays an important role when adopting DevOps or SAFe, and significant mindset changes are needed, which many can resist [13,21]. Many managers resist hybrid models due to "loss of control" over physical ideation processes. Successful transitions require leaders to model agile behaviours, like xFarm's digital empathy rituals [20]. Organisational resistance to Agile principles and a lack of infrastructure support hinder hybrid adoption [14]. Management is sceptical due to the unfamiliarity with Agile and traditional principles [7]. Another obstacle is the transition from hierarchical to decentralised decision-making [8].

Scaled agile is challenging to coordinate, as managing dependencies across Agile Release Trains (ART) and teams is complex due to the need to align Agile teams with traditional project milestones [2,13]. Distributed teams struggle with dependency management, requiring hybrid tools such as Kanban-Gantt integrations [2,11]. For example, a multinational IT project reported delays due to misalignment between Agile sprints and Waterfall milestones [2]. Risks like "handovers due to personnel turnover" and "change requests during sprints" exacerbate coordination challenges, necessitating hybrid workflows [11]. Hybrid methodologies require expertise in both Agile and Traditional methods [19].

Balancing Agile's flexibility with traditional project management's rigidity requires careful integration and robust coordination to succeed [8,10,14]. This is crucial to avoid over-control in adaptive areas, which could still stifle creativity [18]. The prioritisation of mistakes due to poor testing standards in Agile projects underscores the need for hybrid models to enforce compliance without sacrificing agility [11,18]. Many regulated industries like healthcare and construction must blend agile speed with waterfall's audit trails, increasing the administrative burden. For instance, hybrid models in safety-critical industries must maintain rigorous documentation while iterating quickly [2,6]. Documentation and compliance cause a lot of overhead and can burden teams, especially DevOps teams, who face compliance challenges [2,21]. It's more challenging to ensure accurate scope coverage and knowledge retention in the documentation of hybrid/distributed teams [12]. Lack of a common framework for constructing hybrid methods, requiring statistical analysis to identify viable combinations [24]. Frequent, unplanned changes in Agile can disrupt workflows and increase costs if not appropriately managed [14]. It can also be challenging to customise workflows to align with industry-specific needs like compliance versus innovation [8].

### 4.4 Success factors in hybrid project management

Many hybrid implementations are successful and share common strategies. A hybrid method is more likely to be successful if the organisation cooperates and provides leadership support, appreciation, and motivation. Increased customer involvement can increase satisfaction [12]. Identifying statistically validated core practices like code reviews, release planning, and base models like Scrum and Waterfall ensures consistency [24]. Early and continuous involvement of stakeholders ensures alignment with business goals [14]. Hybrid approaches include clear frameworks, stakeholder communication, and adaptive planning, which enhance project management. Additionally, phrased integration and previously mentioned components all lead to a clearly defined, adaptable project, creating a project for a happy, included stakeholder [15].

Hybrid has more stakeholder engagement and better cultural adaptability and leverages Agile's iterative feedback while maintaining traditional governance and compliance [9]. It also includes contextual flexibility, robust documentation, and project managers liaising between Agile teams and organisational hierarchies [16]. Successful hybrid methods are context-specific, combining Agile and traditional elements to fit organisational needs. Empirical process control, such as transparency, inspection, and adaptation, is essential in hybrid scaling. Organisations adapting hybrid models to their specific needs make the model more successful, due to being more adaptable [2]. This empirical control is highlighted in the Agile-Cloud-DevOps integrations study, supporting its importance [21]. For example, xFarm's Slack channels reduced misalignment risks by centralising meeting minutes and client feedback [20].

Furthermore, SAFe's structured framework enables scalable hybrid adoption [13]. Context-specific models like Agile-Stage-Gate or Water-Scrum-Fall are tailored to fit the project's needs, making it more successful [19]. Combining Agile's iterative feedback with traditional documentation and risk management ensures more balanced processes [14]. The introduction of release planning and daily meetings helped reduce delays and balance the delivery tasks with the project plan, improving productivity and collaboration [17]. Leaders also play an essential role and must understand agile and traditional paradigms to guide hybrid transitions. The product owners play a vital role in bridging agile teams and stakeholder expectations [14,18]. Training leaders and team members is essential, as they must navigate both Agile and traditional mindsets to foster collaboration, especially DevOps teams [13,21].

Leaders must master hybrid facilitation. For example, using timeboxed empathy, by dedicating 15-minute check-ins, reinforces psychological safety and boosts team innovation [20]. Skilled teams in hybrid scenarios can self-align and adapt to hybrid workflows, resulting in more positive outcomes [14]. Additionally, trained leaders should implement these hybrid methods, as this will allow teams to operate autonomously with transparency, clear boundaries, and ensure innovation and operational efficiency [18]. Organisations should also invest in cross-functional training and external coaching to navigate these practices, which can lead to better success [7].

Additional success factors and influences include retrospectives and feedback loops that refine hybrid workflows. For example, Scrumban uses Kanban's continuous flow alongside Scrum's iterative reviews to refine processes [2]. DevOps incorporates continuous testing, integration, and monitoring for ongoing improvement [21]. Hybrid retrospectives combining virtual whiteboards and in-person voting sessions improved process adaptation rates, compared to purely digital reviews, a potential field to explore further [22]. SAFE's retrospectives and Inspect & Adapt meetings ensure iterative refinement [13]. Additionally, hybrid teams must prioritise adaptability and learning capability, which enhances project success [12]. Lastly, hybrid methodologies balance flexibility and risk control, reducing the project's risks [19].

## 5  CONCLUSION

This SLR addressed the evolution of Agile into hybrid frameworks and their key challenges/success factors. The shift reflects a pragmatic response to Agile's limitations and increasing project complexity at scale, leading to frameworks like SAFe, Agile-Stage-Gate, and Water-Scrum-Fall that blend adaptability with predictability. Findings show hybrid models address enterprise coordination, compliance, and documentation needs while retaining Agile's iterative feedback and engagement benefits. Successful implementation requires overcoming cultural resistance, coordination complexity, and skill gaps, with leadership buy-in, cross-functional training, and context-specific adaptations being critical. Trends highlight technology (AI, DevOps) enhancing scalability and the need for sector-specific tailoring. Future research should focus on assessing long-term efficacy, especially in regulated sectors (healthcare, finance, construction), and exploring AI's potential to automate workflows and reduce overhead. Hybrid frameworks remain essential for balancing innovation with governance; when implemented properly, they provide the critical balance between agility and control needed in modern project delivery.